\documentstyle[11pt]{article}
\def\be{\begin{equation}}
\def\bear{\begin{array}}
\def\eear{\end{array}}
\def\bea{\begin{eqnarray}}
\def\eea{\end{eqnarray}}
\def\Br{\Biggr}
\def\br{\biggr}
\def\Bl{\Biggl}
\def\bl{\biggl}
\def\l{\label}

\def\ee{\end{equation}}

\def\c{\cite}
\begin{document}
\begin{center}
\LARGE {\bf Non-local 2D Generalized  Yang-Mills theories on
arbitrary  surfaces with boundary}
\end{center}
\begin{center}
 {\bf Kh. Saaidi {\footnote {E-mail: ksaaidi@ipm.ir}}}\\
 {\it Department of Science, University of Kurdistan, Pasdaran Ave., Sanandaj, Iran} \\
 {\it Institute for Studies in Theoretical Physics and Mathemaics,
 P.O.Box, 19395-5531, Tehran, Iran}\\
\end{center}
\vskip 3cm
\begin{center}
{\bf{Abstract}}
\end{center}
 The  non-local generalized  two dimensional
Yang Mills
 theories on  an arbitrary orientable and  non-orientable surfaces
with boundaries is studied. We obtain the effective action of
these theories  for the case which the gauge group is near the
identity, $U\simeq I$. Furthermore, by obtaining the effective
action  at the large-N limit, it is shown that the phase structure
of these theories  is the same as that obtain for these theories
on orientable  and non-orientable surface without boundaries. It
is seen that the $\phi^2$ model of  these
 theories on an arbitrary orientable and  non-orientable surfaces
with boundaries have
 third order phase transition only on $g=0$ and $r=1$ surfaces, with modified area
  $\tilde{A}+{\cal A}/2$ for orientable  and $\bar{A}+\mathcal{A}$ for non-orientable
  surfaces respectivly.

\newpage
{\section{Introduction}}

 The ${\bf BF}$
theory is defined  on a Riemann surface by the Lagrangian $i{\rm
tr }(BF) + {\rm tr}(B^2)$, which is equivalent to the Yang- Mills
theory. This theory has certain properties such as invariance
  under area preserving diffeomorphisms and lack of any
propagating degrees of freedom \c{b1}. These properties are also
shared by a large  class of theories, called the generalized two
dimensional  Yang-Mills (gYM$_2$'s) theories. These theories,
however, are defined  by replacing  an arbitrary class function of
$B$ instead  of tr$(B^2)$ \c{e1}. Several aspect of this theories
such as, partition function, generating functional and large -N
limit  on an arbitrary two dimensional  orientable  and
non-orientable surfaces has been discussed in [11-17]. There is
another way to generalize YM$_2$ and gYM$_2$ and that is to use a
non-local action for the auxiliary field, leading to the so-called
non-local YM$_2$ (nlYM$_2$) and non-local gYM$_2$(nlgYM$_2$)
theories, respectively \c{kh1}.  Several aspects of nlYM$_2$ and
nlgYM$_2$, such as, classical behavior, wave function, partition
function, generating functional, and also large- $N$ limit of it,
have been studied on orientable and non-orientable  surfaces in
[18-21]. In all of these theories, the solution  appear as some
infinite summations over the irreducible representations of the
gauge group. In the large - N limit, however, these summations are
replaced by suitable path integrals over continuous parameters
characterizing the Young tableaux, and saddle-point analysis
shows that the only significant representation is the classical
one, which minimizes some effective action. This continuous
parameters characterizing the representation  is a constrained,
as the length of the rows of the Young tableau is non-increasing.
So for small values of the surface area, the classical solution
satisfies the constraint; for large values of the surface area,
it dose not. Therefore  the dominating representation is not  the
one, which minimizes the effective  action. This introduces a
phase transition  between these  two regime. It can be seen that
the $\phi^2 +{2\alpha \over 3} \phi^3$ and all order of
$\phi^{2k}$ models of nlgYM$_2$ for orientable surfaces with no
boundaries have a third order phase transition on orientable
surfaces with $g\neq 1$ and $g =0$ respectively \c{kh4}. Also,
for non-orientable surfaces without boundary the all order of
$\phi^{2k}$ models of nlgYM$_2$ have a third order phase
transition only on projective plan(RP$^2$), and the $\phi^2
+{2\alpha \over 3} \phi^{2k+1}$  ($k= 0, 1, 2, \ldots $) model of
nlgYM$_2$ on non-orientable surfaces without boundary has the
same phase structure of non-local two dimensional Yang-Mills
theory \c{kh5}. Furthermore, some aspects of Yang-Mills
theory(local and non-local) have been studied for surfaces with
boundary  in [22, 23 24]. The large-N limit of YM$_2$ on cylinder
and on a sphere with three holes have been studied in \c{gr}. The
large gauge group of the non-local YM$_2$ and generalized YM$_2$
on a cylinder have been studied in \c{kh3}, and the authors of
\c{k6} have studied the large-N limit of the two dimensional
$U(N)$ Yang-Mills theory on an arbitrary orientable compact
surface with boundaries. It is seen that, in these cases, for
each boundary, the character if the gauge filed corresponding to
that boundary appears in the expression of the partition function.

The scheme of this paper is the following. In sect.2, I obtain the
partition function of the  non-local generalized  two dimensional
Yang-Mills theory for the case which the gauge group is near the
identity. In sect. 3, I study the large-N limit of nlgYM$_2$ on an
orientable  and non-orientable surfaces with boundaries.  It will
be shown that  the order of phase transition  for $\phi^2$ model,
is 3 with modified area $\tilde{A} + \cal {A}/2$ and $\bar{A} +
\cal {A}$ for orientable  and non-orientable surfaces
respectively.

{\section{The partition function of nlgYM$_2$ on surfaces with
boundaries}}

In \c{kh1}, a non-local generalized  two dimensional Yang-Mills
(nlgYM$_2$'s) theories was defined as:

\be \l{1}
 e^S:= \int DB \exp {\Br \{} \int i{\rm tr}(BF) d\mu +
\omega {\br [}\int \Lambda(B) d\mu {\bl]}{\Bl \}},
 \ee
where B is an auxiliary field at the adjoint representation of
gauge group, F is the field strength,
  $d\mu$ is the invariant measure of the surface;
  $d\mu :={1 \over 2}\epsilon_{\mu \nu} dx^{\mu}dx^{\nu}$ and $\Lambda$ is a
   similarity-invariant function.
It was further shown that the partition function for this theory
for an  arbitrary surface, $\Sigma_{g,n,q}$, with area $A$ and $n$
boundaries  is given by the exact formula \c{{kh1},{kh2}} as:
 \be\l{2}
 Z(U_1,\ldots, U_n;A(\Sigma_{g,n,q}))= \sum_R h_R^q d^{(2-2g-q) }\prod_{i=1}^n
 {\br (}{\chi(U_i) \over d_R}{\bl )}\exp{\br\{}\omega
[-A(\Sigma_{g,n,q})C_\Lambda(R)]{\bl\}}.
 \ee
  Here $\Sigma_{g,n,q}$ is a surface
   containing $g$ handles and $q$ projective planes.
  $R$'s label the irreducible representation of the gauge
group,  $d_R$ is the dimension of the representation $R$,
  $C_\Lambda(R)$ is a linear function of  Casimirs of gauge group in the $R$ representation
   and $\chi_{R}(U)$ is the gauge group character.
    It is clearly
   seen from (2), corresponding to each boundary a factor $\chi_R(U_i) /
   d_R$ appears in  the expression of the partition function.
We  want to obtain  the character of the gauge group for the case
$U \simeq I$. We assume that the gauge group is simple. A group
element is characterized   by some parameters such as: \be U =
e^{\theta^\alpha J_\alpha},\ee where $J_\alpha$'s are the
generators of the gauge group. For $U\simeq I$  and by expanding
the gauge group $U$ up to second order of $\theta^\alpha$'s, one
can arrive at \be U = I + \theta^\alpha J_\alpha + {1\over 2}
\theta^\alpha\theta^\beta J_\alpha J_\beta + \ldots \ee  then by
using of this fact that the representation of the generators of
simple groups are trace less, we obtain \be \chi(U) = d_R + {1
\over 2}\theta^\alpha\theta^\beta \chi_R( J_\alpha J_\beta )+
\ldots \ee in which for a simple group, \be \chi_R( J_\alpha
J_\beta )= {d_R \over d_U}C_2(R){\cal B}_{\alpha\beta} ,\ee where
$d_U$ and ${\cal B}$ are the dimension and the Killing form of the
group, respectively. Therefore for a simple group up to second
order of $\theta^\alpha$'s, we have \be \prod_{j=1}^n{\chi_R(U_j)
\over d_R} = e^{-{{\cal A} \over 2N} C_2(R)}, \ee where \be {\cal
A} = -{ N \over d_U }{\cal
B}_{\alpha\beta}\sum_{j=1}^n\theta_j^\alpha \theta_j^\beta . \ee
So by institute (7) in (2), one arrives at \be\l{2222}
 Z(U_1,\ldots, U_n;A(\Sigma_{g,n,q}))= \sum_R h_R^q d^{(2-2g-q) }\exp{\br\{}
  -{{\cal A} \over 2N}C_2(R) + \omega
[-A(\Sigma_{g,n,q})C_\Lambda(R)]{\bl\}}.
 \ee
Note that the exponent in (9) consists of two parts. The first
part coming from the characters and  the second part depends on
the non-locality term of the action of  the non-local generalized
two dimensional Yang-Mills theory (1).
 Therefore for orientable
   surfaces $q=0$, and the partition function on  an orientable surface
   is
   \be\l{131}
 Z(U_1,\ldots, U_n;A(\Sigma_{g,n})= \sum_R  d^{(2-2g) }\exp{\br\{}
  -{{\cal A} \over 2N}C_2(R) + \omega
[-A(\Sigma_{g,n})C_\Lambda(R)]{\bl\}},
   \ee
    but for non-orientable surface $q\neq 0$ and   $h_R$ is defined as
   \be\l{3}
   h_R := \int\chi_R(U^2)dU.
   \ee
   Here $h_R$ is zero unless the representation $R$  is self- conjugate. In
   this case, the
   representation has an invariant bilinear form. Then, $h_R = +1(-1)$if
   this form is symmetric(antisymmetric), so the
   partition function on non-orientable surfaces is
   \be\l{132}
 Z(U_1,\ldots, U_n;A(\Sigma_{g,n, q})= \sum_{R=\bar{R} } d^{(2-2g-q) }\exp{\br\{}
  -{{\cal A} \over 2N}C_2(R) + \omega
[-A(\Sigma_{g,n,q})C_\Lambda(R)]{\bl\}},
   \ee
where    the summation is only over self-conjugate representation
of the gauge group.

\section{Large-N limit }
\subsection{Large-N limit  on orientable surfaces }
 In the large-N limit of the $U(N)$  gauge group,
 we assume that $C_\Lambda$ is  linear function of
 the rescaled  Casimirs of the gauge group, in which the
 $k$th rescaled Casimir of the gauge group $U(N)$ is as following
\be \tilde{C}_k(R)=: {1 \over N^{k+1}}\sum_{i=1}^N(l_i + N -i)^k,
\ee where $l_i$'s characterize the representation of the gauge
group which satisfy   $ l_i \geq l_j ( i\geq j)$ and it is found
that \be
 d_R = \prod_{1\leq i\leq j\leq N}
{\br (}1 + {l_i-l_j \over j-i} {\bl )}.
 \ee
 One can define a
function $W$ as \be -N^2W\left[ A(\Sigma)\sum_k a_k\tilde
C_k(R)\right] :=w[-AC_\Lambda (R)], \ee  so in the large-$N$
limit, (10) is written as \be\l{11}
 Z = \int D\phi(x)
 \exp{{\Br \{}-N^2{\br (}S_0[\phi] + S_1[\phi]{\bl )}{\Bl \}}}.
 \ee
 where
 \bea
 \phi&:=&{{i-n_i-N}\over
N},\\
 x&:=&{i\over N},
 \eea
and \be S_0[\phi] = -{{\cal A} \over 2}\int_0^1 \phi^2(x)dx, \ee
where is coming from the characters and  boundaries. Also
\begin{equation}\l{12}
S_1[\phi ] = W{\Biggr (}A \int _{0}^{1} G[\phi(x)] dx{\Biggl )} -
(1-g))\int_{0}^{1} dx \int_0^{1} dy \log|\phi(x)- \phi(y)|,
\end{equation}
\be G[\phi(x)]= \sum_k (-1)^k a_k  \phi^2(x). \ee $S_1[\phi(x)]$
is the same action which appear on non-local generalized  two
dimensional Yang- Mills theory on closed surface. It is obvious
that the phase structure of this term is equivalent to the phase
structure  of the following action \c{kh4}
\begin{equation}\l{13}
S'_1[\phi ] = {\tilde{A}} \int _{0}^{1} G[\phi(x)] dx -
(1-g))\int_{0}^{1} dx \int_0^{1} dy \log|\phi(x)- \phi(y)|,
\end{equation}
with
 \be\l{14} \tilde{A} = 2A W'{\Biggr (}A \int _{0}^{1}
G[\phi(x)] dx{\Biggl )}. \ee Therefore we can write the effective
action  which explain the phase structures of this theory as \be
S_{eff}[\phi(x)]= \int_0^1 {\cal L }[\phi(x)] dx
+(g-1))\int_{0}^{1} dx \int_0^{1} dy \log|\phi(x)- \phi(y)|,\ee
where \be {\cal L}[\phi(x)] = {{\cal A} \over 2}\phi^2 +
\tilde{A}G[\phi]. \ee It is remarkable that $S_{eff}$ is nearly
the same  action for the corresponding {\it local } generalized
Yang-Mills Theory. The differences are an additional term which
coming from characters of the gauge group and the existance of
$\tilde{A}$ instead of $A$.

\subsection{Large-N limit on non-orientable surfaces}
Starting from (12), note that the sum in (12) is only over self
conjugate representations. So by applying this additional
constraint  to the sum and then on continuum variable in the large
N limit  and repeat the same approach in the previous subsection,
we obtain  \be\l{33}
 Z = \int D\psi(x)
 \exp{{\Br \{}-N^2{\br (}S_0[\psi] + S_1[\psi]{\bl )}{\Bl \}}},
 \ee
 where the function $\psi(x)$ being defined on the interval $[0,
 1/2]$, in which $\psi(1/2) = 0$, and
 \be
 S_0[\psi(x)] = {\cal A} \int_0^{1\over 2} \psi^2(x) dx,
 \ee which coming from characters of the gauge group and
\begin{equation}\l{34}
S_1[\psi ] = W{\Biggr (}2A \int _{0}^{1/2}G[ \psi (x)] dx{\Biggl
)} -2(1-(g +q/2))\int_{0}^{1/2} dx \int_0^{1/2} dy \log|\psi^2
(x)- \psi^2 (y)|.
\end{equation}
It is obvious that the phase structure of this term is equivalent
to the phase structure  of the following action \c{kh5},
\begin{equation}\l{36}
S'_0[\psi ] = \bar{A} \int _{0}^{1/2} G[\psi (x)] dx -2(1-(g
+q/2))\int_{0}^{1/2} dx \int_0^{1/2} dy \log|\psi^2 (x)- \psi^2
(y)|,
\end{equation}
where $$\bar{A} = 4AW'{\br (} 2A\int _{0}^{1/2} G[\psi (x)] dx{\bl
) }. $$ So the effective action which explain the phase structures
of this theory(nlgYM$_2$)  on a non-orientable surface
 is \be S^{no}_{eff}[\psi]= \int _{0}^{1/2}
{\cal L}^{no}[\psi (x)]dx -2(1-(g +q/2))\int_{0}^{1/2} dx
\int_0^{1/2} dy \log|\psi^2 (x)- \psi^2 (y)|,
\end{equation}
where \be {\cal L}^{no}[\psi (x)] = {\cal A}\psi^2(x) +
\bar{A}G[\psi(x)]. \ee

It is seen that this partition function  is equal to the partition
function on a non-orientable surface with modified  area $\bar{A}
$, genus $g$, $q$ copies of projective plane ($RP^2$), and without
boundaries \c{kh5}. So that  we can obtain the phase structures of
this theory with the same procedure in \c{kh5}.
\section{The  $ G[\phi] = {1 \over 2} \phi^2$ model}
This model defines a non-local Yang-Mills theory. In this case the
effective action on an orientable and non-orientable surfaces are
\be S^{0}_{eff}[\phi(x)]= ({{\cal A} \over 2} + \tilde{A})\int_0^1
\phi^2(x) dx +(g-1))\int_{0}^{1} dx \int_0^{1} dy \log|\phi(x)-
\phi(y)|,\ee  \be S^{no}_{eff}[\psi]= ({\cal A}+ \bar{A})\int
_{0}^{1/2} \psi (x)dx -2(1-(g +q/2))\int_{0}^{1/2} dx \int_0^{1/2}
dy \log|\psi^2 (x)- \psi^2 (y)|, \ee respectively. It is shown
that this theory has third  order phase transition on an
orientable surface with  boundary , $g = 0$ and modified area
$({{\cal A} \over 2} + \tilde{A})$. Also this model has third
order phase transition on non-orientable surface with boundary,
$g=0, q=1$ and modified  area (${\cal A} + \bar{A}$).

\section{conclusion}

I study the  non-local generalized two dimensional $U(N)$
Yang-Mills (nlgYM$_2$) theories on an  arbitrary orientable and
non-orientable surface with boundaries. We obtain the action of
this theory for the case which the holonomies of the gauge field
on boundaries are near the identity, $U\simeq I$, on arbitrary
surface. By obtaining the effective action of these theories at
the large-N limit it is shown that, the effective action of it is
the same as that obtain  on an arbitrary orientable and
non-orientable surface without boundaries with the same genus and
projective plan but with modified area  and an addition term
which coming from characters of the gauge group. It is seen that
the $ \cal{A} $ term  is a function of holonomies of the gauge
field only, in which for orientable and non-orientable surfaces
with the same boundaries is equal. Furthermore, for $G[\phi] =
\phi^{2k +1} (k \in Z)$  the functional ${\cal L}^{no}_{eff} =
{\cal A} \phi^2$. Therefore the phase structures of $G[\phi] =
\phi^{2k +1} (k \in Z)$  models  on non-orientable surface is
equal to the phase structures  of $\phi^2$ model  of local $2D$
Yang-Mills theory  with modified area equal to ${\cal A}$.

\end{document}